# Quorum sensing contributes to activated B cell homeostasis and to prevent autoimmunity.

Caroline Montaudouin[1,2], Marie Anson[1,2], Yi Hao[1,2], Susanne V. Duncker[1,2],  Tahia Fernandez[1,2], Emmanuelle Gaudin[1,2], Michael Ehrenstein[3], William G. Kerr[4], Jean-Hervé Colle[1,2], Pierre Bruhns[5,6], Marc Daëron[5,6], António A. Freitas[1,2]

[1]Institut Pasteur, Départment d'Immunologie, Unité de Biologie des Populations Lymphocytaires, Paris, France, [2]CNRS, URA1961, Paris, France, [3]University College, London, UK, [4]SUNY Upstate Medical University, Syracuse, NY, USA [5]Institut Pasteur, Départment d'Immunologie, Unité d'Allergologie Moléculaire et Céllulaire, Paris, France [6]Inserm, U760, Paris, France.

**Running title:** Quorum sensing in B cell homeostasis.

Correspondence to: Antonio A. Freitas, UBPL, Institut Pasteur, 25 rue du Dr. Roux, 75015 Paris. Tel: +33145688582; Fax: +33145688921; email: antonio.freitas@pasteur.fr

**1**

---

[1] [1]This work was supported by grants from the European Research Council (ERC), the Agence Nationale pour la Recherche (ANR) and the Agence pour la Recherche sur le Cancer (ARC) and by the Institut Pasteur, Centre National pour la Recherche Scientifique (CNRS) and Institut National de la Santé et de la Recherche Médicale (INSERM). CM, YH, and EG were supported by the Direction Recherche Enseignement et Technologie (DRET), TF by the Fondation IMABIS (Spain) and the Fondation de le Recherche Médicale (FRM) and WGK by the U.S. NIH (RO1 HL72523, R01 101748) and the Paige Arnold Butterfly Run.

**Abstract**

Maintenance of plasma IgM levels is critical for immune system function and homeostasis in humans and mice. However, the mechanisms that control homeostasis of the activated IgM-secreting B cells are unknown. After adoptive transfer into immune-deficient hosts, B-lymphocytes expand poorly but fully reconstitute the pool of natural IgM-secreting cells and circulating IgM levels. By using sequential cell transfers and B cell populations from several mutant mice, we were able to identify novel mechanisms regulating the size of the IgM-secreting B cell pool. Contrary to previous mechanisms described regulating homeostasis, which involve competition for the same niche by cells having overlapping survival requirements, homeostasis of the innate IgM-secreting B cell pool is also achieved when B cells populations are able to monitor the number of activated B cells by detecting their secreted products. Notably, B cell populations are able to assess the density of activated B cells by sensing their secreted IgG. This process involves the FcγRIIB, a low-affinity IgG receptor that is expressed on B cells and acts as a negative regulator of B cell activation, and its intracellular effector the inositol phosphatase SHIP. As a result of the engagement of this inhibitory pathway the number of activated IgM-secreting B cells is kept under control. We hypothesize that malfunction of this quorum-sensing mechanism may lead to uncontrolled B cell activation and autoimmunity.

**Introduction**

Maintenance of plasma IgM levels is critical for innate and adaptive immune system function and homeostasis in humans and mice. Decreased IgM levels result in diminished innate protection against bacterial invasion (1, 2). In humans, splenectomy, a therapeutic measure following trauma, cancer or autoimmune diseases, results in increased susceptibility to bacterial infections (3); asplenic (Hox-11$^{-/-}$) or splenectomized mice also show diminished immune responses to bacterial infection (2). Such reduction in the innate protection against bacterial invasion is related to decreases in both natural plasma IgM levels and the number of IgM-secreting cells (2). In contrast, increased IgM titers are often associated with autoimmunity (4, 5). For example, humans develop autoimmune disorders when they have primary immune deficiencies that are characterized by elevated natural IgM levels when B-lymphocytes are unable to switch to IgG producing cells. These deficiencies could be due either to defective T-B cell cooperation, as is the case of CD40/CD40L deficiencies, or to an intrinsic inability of the B cells to perform class switch recombination (4, 5). Similarly, increased IgM levels in mice are associated with several autoimmune disorders (6). For example, mice with activation-induced cytidine deaminase (AID) defects that prevent class switch recombination develop hyper IgM-like syndromes that are associated with autoimmune diseases (7). Thus, any failure to maintain the homeostasis of IgM-secreting B cells is deleterious, either by increasing susceptibility to infection or by inducing autoimmune disorders; however, the underlying mechanisms are unknown. In most cases, homeostatic regulation is achieved by competition of different cells for the same "survival niches" (8, 9). However, it is unclear whether such mechanism would also apply to IgM homeostasis, as not only secreting B cell numbers should be maintained, but also the total amount of IgM they produce must also be regulated. It is also not clear, which mechanisms could limit lymphocyte numbers during immune responses, in situations where resources are not limiting and physiological niches may be disrupted.

In here we report an experimental strategy allowing studying the mechanisms involved in this regulation. Indeed, in contrast to T lymphocytes that undergo considerable homeostatic expansion when transferred into immune-deficient (10) naïve B lymphocytes expand poorly but reach a stable equilibrium and fully reconstitute the pool of IgM-secreting B cells and circulating IgM levels (11). The size and composition of this pool is tightly controlled as it remains stable for up to 6 months after transfer, independently of the number of injected naïve B cells (11). Thus, in these host mice the number of Ig-secreting cells is kept under strict homeostatic regulation as observed in intact mice (8). This model is therefore ideal to study the mechanisms of homeostasis of the number of IgM-secreting B cells, since only an adoptive transfer strategy allows the follow-up of different B cell populations (recognized by different allotypes) in the same mouse. Indeed, by using sequential transfers of B cell populations from several mutant mice, we identified feedback mechanisms regulating the size of the IgM-secreting B cell pool in a B-cell specific manner, i.e. excluding side effects induced by the mutations in other non-lymphoid cells. We found that contrary to the previous mechanisms described regulating homeostasis, which involve competition for the same niche by cells sharing overlapping survival signals (8, 9), homeostasis of the innate IgM-secreting B cell pool is also achieved when B cell populations are able to monitor the number of activated B cells by detecting their secreted products. Notably, B cell populations are able to assess their density and limit the number of activated IgM-secreting B cells when they sense the levels of secreted IgG via FcγRIIB, a low affinity IgG receptor that is expressed on B cells and acts as a negative regulator of B cell activation (12) by a SH2-containing inositol-5-phosphatase 1 (SHIP1)-mediated pathway (12, 13). These results reveal a new mechanism of homeostatic regulation, which recalls mechanisms found in procariots, which have been collectively named as quorum sensing (14, 15). Importantly, it finally explains the development of autoimmune conditions when IgG production is impaired, and the apparent paradox of the beneficial effects of IV Ig therapy in several autoimmune disorders (16).

**Materials and Methods:**

**Mice and cell transfers.** C57BL/6(B6).Ly5$^b$IgH$^b$, B6.Ly5$^a$IgH$^b$, B6.Ly5$^b$IgH$^a$, B6.Ly5$^a$IgH$^a$, B6.Ly5$^a$IgH$^a$IgM$_s^{-/-}$ (17), B6.Ly5$^b$IgH$^b$CD3$\epsilon^{-/-}$, B6.Ly5$^a$IgH$^a$CD3$\epsilon^{-/-}$, B6.Ly5$^b$IgH$^b$Fc$\gamma$RIIB$^{-/-}$ (12), B6.Ly5$^a$IgH$^b$SHIP1$^{-/-}$ (18), and B6.Rag2$^{-/-}$ mice were purchased from Charles River, France, ;  the Centre Des Techniques Avancées (CDTA), Centre Nationale de la Recherche Scientifique (CNRS), Orleans, France, ; or Taconic, Hudson, NY, USA; or were from our animal facilities at the Institut Pasteur. LN cells containing approximately 5x10$^6$ B cells were injected i.v. into Rag$^{-/-}$ hosts. In experiments involving sequential cell transfers the injected populations differed by allotype markers (Ly5$^{a\ or\ b}$ and IgH$^{a\ or\ b}$ allotype). Recipient mice were bled and killed at different times after cell transfer. The number of B cells derived from each cell population was calculated. Experiments were preformed according to the Pasteur Institute Safety Committee in accordance with French and European guidelines and the ethics Committee of Paris 1 (permits 2010-0002, -0003 and -0004).

**Flow Cytometry and cell sorting.** Cells were stained using anti-CD19 (1D3), anti-IgM$^a$ (DS1), anti-IgM$^b$ (AF6-78), anti anti-Ly5$^a$ (A20), anti-Ly5$^b$ (104), anti-CD21 (7G6), anti-CD23 (B3B4), anti-IgM (R6-60.2), and anti-IgD (11-26c.2a) monoclonal antibodies from BD Pharmingen coupled with the appropriate fluorochromes. Four/six color staining used the appropriate combinations of FITC, PE, TRI-color, PerCP, PECy7, biotin, APC, AlexaFluor647 and APCCy7-coupled antibodies. Biotin-coupled antibodies were secondarily labeled with APC-, TRI-Color- (Caltag, San Francisco, CA, USA), PerCP- (Becton Dickinson, San Jose, CA, USA) or APCCy7-coupled (Pharmingen) streptavidin. Dead cells were excluded based on light scattering. All acquisitions and data analysis were performed with a FACSCanto or LSRFortessa (Becton Dickinson, San Jose, CA, USA) interfaced to with the Macintosh FlowJo software. Subsets of follicular (CD21$^{high}$CD23$^{high}$), marginal (CD21$^{high}$CD23$^{low}$) or activated (CD21$^{low}$CD23$^{low}$) B cells were sorted on a FACSAria (Becton Dickinson, San Jose, CA, USA) cytometer. The purity of the sorted populations varied from 90-95%.

**ELISA and ELISA Spot Essay.** Sera Ig concentrations were quantified by ELISA. Plates were coated either with antibodies to total IgM, IgM[a] or IgM[b] and saturated with PBS-1% gelatin. Dilutions of sera were added. After incubation (1 hour, 37°C) and washing, PO-labeled goat anti-mouse IgM antibodies were added. After incubation and washing, bound antibodies were revealed with the substrate O-phenylenediamine and $H_2O_2$. The reaction was stopped after 10 min. by addition of 10% SDS and the absorbance read at 450nm in a titertek multiscan spectrometer (Flow laboratories, Irvine, Scotland). Titration of serum IgM and IgG was performed using purified mouse IgM or IgG as standards purified mouse IgM or IgG (Southern Biotechnology, Inc.). Ig concentrations were determined by comparing the displacement of the dilution curves in the linear interval between standards at a concentration of 10 mg/ml and the serum samples. Specificity of the reactions for the detection of both IgM[a] and IgM[b] was confirmed using serum from B6.IgH[a] and B6.IgH[b] donor mice. For autoantibodies, plates were coated with mice mouse tubulin or DNA. To test for OxLDL, we used the OxLDL-Ab ELISA kit purchased from IMTEC. The quantification of IgM secreting cells was assayed by ELISA Spot Essay technique. Briefly, plates were coated with goat anti-mouse IgM antibodies or anti-allotypic IgM antibodies. After saturating, the cells were distributed into the micro wells in RPMI1640-2%FCS. The plates were incubated for 4-6 h at 37°C, 5% $CO_2$ atmosphere. After extensive wash, plates were incubated with goat anti-mouse IgM labeled with alkaline phosphatase. After washing, the revealing substrate was added (2,3mM 5-bromo-4-chloro-3-indolyl phosphate diluted in 2-amino-2-methyl-1-proprenolol buffer).

**Results**

**Replenishment of the IgM-secreting B cell pool is strictly controlled.** Upon adoptive transfer of LN cells from immune competent into immune deficient mice, T cells expanded to reconstitute a considerable fraction of the peripheral T cell pool (10). In contrast, the majority of the $IgM^{low}IgD^{high}CD21^{int/high}CD23^{high}$ follicular LN B cells (Fig. 1a) did not proliferate and therefore, the number of recovered B cells was limited and did not exceed $2 \times 10^6$ cells (11), i.e. less than the number of injected cells. The size and composition of the pool of persisting cells was regulated independently of the number of injected B cells and remained stable for up to 6 months after transfer (11). Most of the persisting B cells expressed activated $IgM^{high}IgD^{low}CD21^{high}CD23^{low}$ marginal zone (MZ) or $IgM^{low}IgD^{low}CD21^{low}CD23^{low}$ phenotypes (Fig. 1a). In the spleen, Ig-secreting cells were recovered with increased frequency, primarily among the $CD21^{low}CD23^{low}$ B cells (Fig. 1b). However, since quantification of the total number of activated B cells in the spleen is not a reliable method to evaluate the total size of the activated B cell pool, as activated B cells could have migrated to organs (as the liver, gut or the flat bones bone-marrow) where numbers cannot be accurately measured, we chose the serum IgM levels as the most reliable indicator of the state of B cell activation and of the total number of IgM-secreting B cells in the whole mouse (Supplementary Note 1). In these host mice we found that the serum IgM and IgG concentrations were similar to those of control mice, despite reduced B cell numbers (Fig. 1c). Interestingly, the titers of self-reactive antibodies were increased compared to donor mice (Fig. 1d), suggesting that the persisting B cell populations were naturally selected on the basis of their reactivity against self and environmental antigens. These findings indicate that the transfer of mature B cells restored a stable and tightly controlled pool of IgM-secreting B cells in immune-deficient mice, but they do not provide insight into the homeostatic mechanisms that limits their number.

**Feedback mechanisms control newly arriving B cell activation and IgM production**. To identify the mechanisms that control the number of the IgM-secreting cells, we investigated whether an established population of B cells could modify the fate of a newly transferred B cell population. To do this, we used a sequential cell transfer strategy as follows. Rag-deficient murine hosts were first injected with LN cells from Ly5$^a$IgH$^a$ donors; second, 4 weeks later, the hosts were injected with LN cells from double congenic Ly5$^b$IgH$^b$ mice (Supplementary Fig. 1). We evaluated the number, phenotype, and function of the B cells recovered 6 weeks after the transfer of the second population into the hosts. We found that, amongst the second population, the number of naïve B cells recovered was not significantly altered by the presence of a first B cell population. However, the fraction and number of activated B cells (CD21$^{high}$CD23$^{low}$ and CD21$^{low}$CD23$^{low}$) was significantly reduced in hosts injected first with B cells from Ly5$^a$IgH$^a$ donors compared to controls (Fig. 2a). Consequently, the amount of IgM$^b$ that the second B cell population generated in the Rag-deficient hosts was significantly lower than the amount generated by the same cell population when transferred alone into naïve hosts (Fig. 2b). At the same time, total IgM levels were slightly greater in mice receiving both B cell populations (Fig. 2b). In summary, we found that in mice injected sequentially with two LN cell populations, B cells from the first population replenished the compartment of activated IgM-secreting cells leading to normal serum Ig levels and used a "feedback" mechanism to regulate activation of the second population of transferred B cells and, ultimately, their IgM production. Thus, by using this sequential cell transfer strategy we unraveled feedback mechanisms that maintained the number of the activated IgM-secreting B cells stable. This feedback regulation could result from competition for cellular niches (8, 9), but could also occur in response to soluble factors secreted by the first B cell population.

**Plasma IgG regulates the number of activated B cell and IgM production.** To investigate whether secreted specific B cell products, such as immunoglobulin, were involved in the feedback regulation mechanism, we investigated the role of IgM produced by the first B cell population. Rag-deficient mice were injected with a population of Ly5[b]IgH[a] LN cells from IgMs[-/-] mice; the B cells of these donor mice express but cannot secrete IgM (17). Four weeks later, mice were injected with LN cells from wild-type (WT) Ly5[a]IgH[b] mice. As a control, the second WT cell population was also transferred into hosts that were not injected with the first cell population. The two groups of mice were sacrificed 6 weeks after transfer of the WT cell population. As shown in Fig. 3a the amount of IgM[b] produced by the WT B cell population were markedly lower in mice already hosting cells from IgMs[-/-] donors than in host mice that received only the WT population (Fig. 3a). We concluded that IgM produced by the first population of B cells was not responsible for the feedback effect, since secretion of IgM was disabled in those cells.

Since B cells from IgMs[-/-] donors can secrete IgG (17) (Supplementary Figure 2), we next investigated the role of IgG in feedback inhibition. B cell class-switch recombination and IgG production requires T cell help and consequently T cell deficient mice show severely reduced levels of serum IgG (19). Therefore, to prevent strong IgG production by the first population of transferred B cells, Rag-deficient mice were injected with LN cells from Ly5[b]IgH[a] CD3ε[-/-] T cell deficient mice (Supplementary Note 2) and, 4 weeks later, with LN cells from Ly5[a]IgH[b] WT donors. Of note, B cells from CD3ε[-/-] mice behave similarly as B cells from WT donors in that the number of recovered cells was identical. We monitored the number of activated B cells and the IgM[b] produced by the second WT B cell population 6 weeks after the second cell transfer. In the quasi absence of IgG secreted by the first population of B cells (Supplementary Figure 2), there was no inhibition of IgM production by the second B cell population (Fig. 3b).

We had observed that the level of feedback inhibition of the IgM production by the second B cell population was variable in different host mice injected with two WT populations. Considering that the feedback inhibition seems to be determined by the presence of IgG, we compared the IgM levels produced by the second B cell populations with the IgG levels present in individual mice from several different experiments. Notably, we found that the amount of IgM produced by the second B cell population inversely correlated with IgG levels in the host serum (Fig. 3c).

To confirm the role of IgG in the feedback regulation of IgM production, we transferred LN cells into Rag2-deficient hosts that were either untreated (control) or injected with purified mouse IgG. The number of B cells with an activated phenotype and the IgM production were significantly reduced in the mice injected weekly with an IgG dose required to reach and maintain the physiological levels of serum IgG (Fig. 3d), but not in mice injected with lower doses of IgG or in mice receiving as control soluble IgMs (Supplementary Note 3 and Supplementary Figure 3). These findings also indicate that the IgG-mediated inhibitory signals require threshold levels of circulating IgGs. If these levels are not attained, as it was the case in mice receiving lower doses of IgG or injected with B cells from CD3[-/-] origin, B cell activation and IgM production are not inhibited.

In conclusion, our findings, clearly demonstrate that an active IgG-dependent suppressive mechanism can regulate IgM production. In the absence of IgG secretion by the first B cell population, IgM production by the second B cell population was not inhibited, and the amount of IgM produced by the second B cell population was inversely correlated with the IgG level in the host serum. More importantly, they indicate a presence of a yet non-described mechanism of homeostatic control. It has been proposed that homeostasis of Ig-secreting cells is due to competition for a common niche of restricted size (8, 9). We now show that, in the absence of competing populations, passively administered IgG reduced activation and further IgM production by subsequently introduced B cells.

**The inhibitory FcγRIIB is required for IgG-mediated feedback regulation**. To investigate the mechanism of action underlying the IgG-dependent control of IgM production, we first injected Rag-deficient mice with LN cells from Ly5[a]IgH[a] WT mice; 4 weeks later, we injected them with a second population of LN cells from either WT or FcγRIIB[-/-] Ly5[b]IgH[b] donors. FcγRIIB is a low-affinity IgG receptor expressed on B cells that acts as a negative regulator of B cell activation (12). We found that the first (WT) B cell population did not suppress the second B cell population that lacked FcγRIIB (Fig. 4a, b), but was able to suppress 50% of the IgM production by a population of WT B cells (Fig. 4b). Inhibition of B cell activation may require FcγRIIB cross-linking, which can be achieved in the presence of either immune complexes or Ig aggregates (20). Notably, as mentioned we found higher titers of self-reactive antibodies in the sera of cell transfer host mice than in donor mice (Fig. 1d). These autoantibodies might form immune complexes that could either cross-link FcγRIIB or co-aggregate B cell receptors (BCR) with FcγRIIB and generate negative signals (20) by enabling tyrosyl-phosphorylation of FcγRIIB and recruitment of SHIP (13). To test this hypothesis, we injected Rag-deficient mice first with Ly5[b]IgH[a] WT cells and then, 4 weeks later, with a population of cells from either WT or SHIP1[-/-] Ly5[a]IgH[b] donors (18). We found that WT B cells, while inhibiting a second population of WT B cells, failed to reduce the number of activated SHIP1[-/-] Ly5[a] B cells or their IgM[b] production (Fig. 4c, 4d). Since these findings were obtained upon adoptive cell transfer into lymphocpenic hosts their validity could be questioned, as the new environment would not reflect that of an intact mouse. Nevertheless, the size and composition of the IgM-secreting B cell pool remains stable for up to 6 months after transfer indicating that its homeostasis mimics the rules observed in intact mice. By using the strategy of sequential transfer of B cell populations from several mutant mice, we were able to study the mechanisms regulating the size of a stable IgM-secreting B cell pool. Notably, our findings obtained using adoptive cell transfers, were fully supported by the observation that both intact non-manipulated $Fc\gamma RIIB^{-/-}$ and $SHIP1^{-/-}$ mice had IgM serum levels that were significantly higher than those of control littermates (Fig. 4e). Additionally, the cell transfer strategy demonstrated that these defects are B-cell specific and not induced by the lack of expression of the FcRIIB or SHIP1 on other non-lymphoid host cells. Taken together, these results show that there are feedback mechanisms that regulate the number of activated B cell and plasma IgM levels and that these mechanisms involve plasma IgG- and $Fc\gamma RIIB$-mediated, SHIP1-dependent negative signals (Supplementary Note 4 and Supplementary Fig. 4).

**Discussion**

The strategies to approach autoimmune disorders displaying elevated immunoglobulin (Ig) levels in the absence of obvious immunization are yet unclear. These disorders include hyper IgM syndromes and other pathological conditions such as SLE. In some cases, therapeutic intervention with the administration of IV Igs were shown to have beneficial effects (16) but it appears paradoxical that in steady-state conditions where Ig levels are already increased and induce pathology, further increases in circulating Ig could still have any beneficial effects. Above and beyond the possible anti-inflammatory effects of the administered Igs (21), the present findings showing that there are feedback mechanisms that regulate the number of activated B cell and plasma IgM levels and that these mechanisms involve plasma IgG- and FcγRIIB-mediated, SHIP1-dependent negative signals, may provide another explanation to this phenomenon.

It has been long known that passively administered antigen-specific IgG can suppress or enhance specific primary thymus-dependent (22, 23) and thymus-independent IgM antibody responses (24). It is generally accepted that IgG-mediated suppression is antigen-specific (22). The suppressive effects could either be due to epitope masking and antigen sequestration by the high affinity IgG preventing initiation of new IgM responses (23), or be mediated by an Fcγ-dependent regulation of B cell activation (for review see (22)). Trials on whether F(ab')2 fragments kept the suppressive activity of intact IgG have given contradictory results (25). It has also been reported that IgG suppressed specific antibody responses as efficiently in FcγR-deficient mice as in wild type controls (26, 27) questioning the role of this inhibitory receptor in the control of immune responses. However, FcγRIIB-deficient mice down regulate late responses to immune complexes (28) or during inflammation (29, 30) suggesting that these receptors may act to prevent overextended immune responses (25). It should be pointed out that studies that investigated the *in vivo* role of IgG-mediated regulation of B cell activation and immune responses have all used actively immunized mice, under conditions where high titers of antigen-specific antibodies were produced (22, 25). The novelty of our approach resides on the fact that we preformed our studies in the absence of any antigenic challenge. In these novel conditions our results show that circulating IgG, constitutively present in non-immunized mice, may have a fundamental role in controlling natural activated B cell numbers and the IgM they produce. They seal a hiatus and provide new valuable information as they demonstrate that FcγRIIB-dependent negative regulation controls B cell activation without any intentional immunization.

We show that regulation of IgM production engages a FcγRIIB-mediated, SHIP1-dependent inhibitory feedback loop in which B cells themselves sense the number of activated B cells by detecting their secreted IgG. When there are enough activated B cells, the high titers of IgG and self-reactive antibodies that they produce can lead either to FcγRIIB-mediated aggregation by immune complexes, possibly inducing activated B cell apoptosis (31, 32) or to co-aggregation of BCR and FcγRIIB, which dampens B cell activation (20). Either mechanism may prevent the accumulation of activated IgM-secreting B cells. However, while direct FcγRIIB cross-linking may inhibit B cells independently of their antigenic-specificity and affect a large fraction of the cells, BCR and FcγRIIB co-aggregation will preferentially inhibit self-reactive B cells. Considering the stronger effects observed using SHIP1-deficient B cells, it is possible that alternative FcγRIIB-independent inhibitory pathways may also participate in the control of the IgM-secreting cell pool. SHIP was indeed shown to control activation/proliferation signals triggered by a variety of receptors that activate PI3 Kinase (13). Once IgG levels reestablished normal B cell activation can proceed in response to new antigenic stimulation.

Our current findings may provide an evolutionary explanation for the expression of inhibitory Fcγ receptors by B cells: By simultaneously allowing detection of plasma IgG levels and inhibiting B cell activation, these receptors contribute to control the number of activated Ig-secreting B cells and prevent autoimmune disorders. They indicate that self-reactive B cells can expand and secrete auto-reactive Igs if not held in check by sufficient IgG in their environment. This notion is supported by other independent data. The inability of B cells to detect plasma IgG because of defects in FcγRIIB expression or signaling leads to increased IgM levels and autoimmune disease (12, 29, 33-35). Altered human FcγRIIB signaling in B cells resulting from the $I_{232}T$ polymorphism (36, 37) as well as decreased FcγRIIB expression in memory B cells (38) have been reported in SLE patients. Inversely, partial restoration of FcγRIIB expression in the B cells of FcγRIIB$^{-/-}$ mice (39) or FcγRIIB overexpression in B cells reduces systemic lupus erythematous (SLE) incidence (40). Finally, the failure to produce IgG either because of early thymectomy in toads (41), inappropriate T-B cell cooperation humans (5) (Supplementary Note 5) or because of the intrinsic inability of AID-mutant B cells to switch to IgG production in both mice and human (4, 7) results in hyper-IgM syndromes and autoimmune pathology. They may also be of relevance to the perception of the proposed "hygiene hypothesis" (42). Indeed, increase incidence of infections in the developing countries may lead to higher serum titers of IgG and thus, help to prevent development of autoimmune and allergic diseases.

Our findings also suggest an as-yet unsuspected physiological role for natural self-reactive antibodies. Specifically, it has never been clear why natural self-reactive Abs should be allowed to exist; but our observation of IgG-mediated feedback regulation of innate IgM in normal non-immunized mice suggests that natural self-reactive antibodies, which make up a significant fraction of the natural plasma Ig (43), by facilitating FcγRIIB aggregation provide a "warning signal" that reflects B cell density. Thus, natural self-reactive Abs might participate in activated B cell homeostasis under physiological steady-state conditions.

In conclusion, these results identify a new mechanism of homeostatic control. So far, in mammals, homeostatic control was believed to be due to competition of cellular populations for a common niche of restricted size, defined by the ensemble of cellular interactions and trophic factors required for cell survival (8, 44). However, in situations where resources are not limiting, i.e. immune responses, excess of self-antigens or cytokines, it is not clear, which mechanisms could limit expanding lymphocyte numbers. Thus, the question: how lymphocyte populations "count" the number of their individuals and how do they "know" when to stop growing? We now show that in absence of competing cell populations, passively administered IgG reduced activation and IgM production by subsequently introduced B cells. Thus, we propose that control of lymphocyte numbers could also be achieved by the ability of lymphocytes to perceive the density of their own populations (44). Such mechanism would be reminiscent of the primordial "quorum-sensing" systems used by some bacteria in which a bacterium senses the accumulation of bacterial signaling metabolites secreted (by the same or other cells), allowing the bacterium to sense the number of cells present in a population and adapt their growth accordingly (14, 15). In summary, these "quorum-sensing" mechanisms allow bacteria to coordinate their gene expression according to the density of their population (14, 15). Quorum sensing can play a critical role in lymphocyte homeostasis with the proviso that lymphocytes have the capacity to assess the number of molecules they interact with and can mount a standard response once a threshold number of molecules is detected. The situation described here seems to support this hypothesis. We found that homeostasis of the innate IgM-secreting B cell pool is achieved when total B cell populations are able to monitor the number of activated B cells by detecting products secreted by some of their members (Fig. 5a). The ability of B cells to detect plasma IgG levels through such a "quorum-sensing" mechanism may be crucial to immune system homeostasis, providing a critical checkpoint mechanism to prevent excessive B cell activation and autoimmunity. Malfunction of this "quorum-sensing" mechanism may lead to uncontrolled B cell activation and autoimmune disease (Fig. 5b, c). It is possible that similar quorum-sensing mechanisms may work to maintain the homeostasis of other lymphocyte populations (44) or control organ size during development.

**Acknowledgements** We thank Drs. B. Rocha for critical assessment of the manuscript and G. Laval for help with statistics. We also thank M-P Mailhé for technical assistance and the Plateforme de Cytométrie de Flux for cell sorting.

**Authorship Contributions:** CM, YH, SVD, MA, TF and EG performed research and analyzed data. ME, WGK, JHC and PB contributed vital new reagents or analytical tools. All of the authors analyzed the results and commented on the manuscript. PB helped to write the paper. MD wrote the paper. AAF designed research, analyzed data and wrote the paper.

**Disclosure of Conflicts of Interest:** The authors have no conflict of interest to declare

**Figure Legends:**

**Figure 1 – Fate of LN B cells transferred into Rag-deficient hosts. a.,** LN cells containing about $5 \times 10^6$ B cells were transferred i.v. into immune deficient hosts. Left dot plots show the IgM, IgD (top), and CD21, CD23 (bottom) phenotype of donor LN B cells before transfer (gated on CD19$^+$ cells). Right dot plots show the IgM, IgD (top), and CD21, CD23 (bottom) phenotype of the donor B cells recovered from the spleen of the host mice 8 weeks after transfer. The relative representation of the different CD21, CD23 subpopulations is displayed. Note that the majority of the B cells recovered 8 weeks after transfer into Rag-deficient hosts showed IgM$^{high}$IgD$^{low}$ or IgM$^{low}$IgD$^{low}$ and CD21$^{high}$CD23$^{low}$ or CD21$^{low}$CD23$^{low}$ activated phenotypes. **b.,** Different populations of B cells follicular (FO - CD21$^{int/high}$CD23$^{high}$), marginal zone (MZ - CD21$^{high}$CD23$^{low}$)**,** and CD21$^{low}$CD23$^{low}$ B cells recovered from the spleen of host mice 8 weeks after B cell transfer were purified by cell sorting and the number of IgM-secreting cells in each B cell subset detected by ELISPOT. Bars show the mean % of IgM-secreting cells among donor B6 spleen B cells and total host spleen B cells as well as of the sorted follicular (FO - CD21$^{int/high}$CD23$^{high}$), marginal zone (MZ - CD21$^{high}$CD23$^{low}$)**,** and CD21$^{low}$CD23$^{low}$ B cells recovered from the spleen of host mice 8 weeks after B cell transfer. Note that the majority of IgM-secreting cells detected by ELISPOT were recovered among the activated CD21$^{low}$CD23$^{low}$ B cells. Each column represents the mean of values for 3 individual mice. Similar findings were obtained in 3 independent experiments. **c.,** Shows the IgM (left) and IgG (right) concentrations in the serum of donor B6 mice and of Rag$^{-/-}$ hosts 8 weeks after B cell transfer. Each bar represents the mean ± sd of values for 8 individual mice. Note that in spite of the low number of B cells recovered, Ig concentrations in these hosts were identical or slightly higher than those present in the donor mice. Similar results were obtained in more than 7 independent experiments. **d.,** Titers (expressed in arbitrary units) of self-reactive anti-mouse DNA (left), anti-mouse tubulin (middle), and anti-mouse OxLDL in the serum of individual donor B6 mice or in the serum of Rag-deficient hosts. The hosts were injected 8 weeks previously with LN B cells from WT B6 mice. Bars indicate the mean values. Note the increased titers of self-reactive antibodies in the serum of the host mice. Statistically significant differences are indicated (* $p < 0.05$; *** $p < 0.001$).

**Figure 2 – Sequential B cell transfers. a.,** Shows the number of FO - $CD21^{int/high}CD23^{high}$ (left panel), MZ - $CD21^{high}CD23^{low}$ (middle panel), and $CD21^{low}CD23^{low}$ (right panel) recovered from the second $Ly5^bIgH^b$ population when injected alone (left) or into mice injected 4 weeks before with a first $Ly5^a$ $IgH^a$ cell population. Each bar represents the mean ± sd of values for 9 individual mice. Note that while the number of resting B cells was not altered, the number of activated $CD21^{high}CD23^{low}$ and $CD21^{low}CD23^{low}$ present among the second set of B cells was significantly lower when these cells are were transferred into host mice homing already injected with a first set of B cells. **b.,** Shows the quantity of $IgM^b$ produced by the second B cell population when transferred into naïve hosts (left) or into a host containing a first B cell population. Each bar represents the mean ± sd of values for 9 individual mice. Note the significantly reduced IgM production by the second B cell population when these cells were transferred into mice containing a first set of B cells transferred 4 weeks before (right). Note that the total IgM concentrations were slightly greater in mice injected with two B cell populations than in the mice injected with the first population only (left). Similar results were obtained in five independent experiments or when the order of injection of the two B cell populations was inverted. Statistically significant differences are shown (* $p < 0.05$; *** $p < 0.001$).

**Figure 3 – Role of immunoglobulin in feedback regulation of IgM production. a,** The $IgM^b$ produced by a second $Ly5^aIgH^b$ B cell population transferred alone into naïve mice or into mice pre-injected 4 weeks before with a first population of cells from $Ly5^bIgH^a$ $\mu s^{-/-}$ donor mice. Each bar represents the mean ± sd of values for 6 mice. Note that in spite of the inability of the first B cell population to secrete IgM, the IgM production by the second population was significantly reduced in these hosts as compared to its production in naïve hosts. Similar results were obtained in a second independent experiment. **b,** The IgM[b] produced by a second Ly5[a]IgH[b] B cell population from normal donors transferred alone into naïve mice or into mice pre-injected 4 weeks before with a first population from Ly5[b]IgH[a] CD3ε[-/-] donors. Each bar represents the mean ± SD of values for 6-9 mice. Note that in this case, the levels of IgM produced by the second B cell population were identical in the two groups of host mice ($p$ = NS). Similar results were obtained in a second independent experiment. **c,** Correlation of the levels of IgM produced by the second population with IgG levels present in the sera of 24 host mice pooled from different experiments. The correlation coefficient is shown ($r^2$ = 0.47, $p$ < 0.001). **d,** The IgM produced by the B cells of a LN population transferred alone into untreated naïve mice or into mice injected i.v. with 1mg of purified soluble mouse IgG (Innovative Research, Inc.) the day before cell transfer and thereafter intraperitoneally every week for 5 weeks. Each bar represents the mean ± SD of values for 10 mice. Note that in mice injected with IgG, the IgM production by the transferred B cells was significantly reduced. Similar results were obtained in a second independent experiment. Statistically significant differences are shown (** $p$ < 0.01; *** $p$ < 0.001).

**Figure 4 – Role of the FcRγIIB and SHIP1 in feedback regulation of IgM production. a,** The number of MZ - CD21[high]CD23[low] (left panel) and CD21[low]CD23[low] (right panel) B cells recovered from the second FcRγIIB[-/-] Ly5[b]IgH[b] cell population when injected alone or into mice injected 4 weeks before with a first Ly5[a]IgH[a] WT population. Each bar represents the mean ± SD of values for 12-14 mice. Note that the number of activated CD21[high]CD23[low] and CD21[low]CD23[low] present among the second set of B cells was not modified by the presence of a first set of B cells. **b,** The IgM[b] titers produced by a second population of WT (left) or FcRγIIB-deficient (right) B cells transferred into Rag-deficient mice injected 4 weeks before

with cells from WT IgH$^a$ donors expressed as a % of the control levels obtained after their transfer into naïve Rag-deficient hosts. Note that the IgM$^b$ production by FcR$\gamma$IIB-deficient B cells was not inhibited in the presence of a first population of WT B cells. Each bar represents the mean ± SD of values for 12-14 mice. Data shown were pooled from two independent experiments. **c,** The number of MZ - CD21$^{high}$CD23$^{low}$ (left panel) and CD21$^{low}$CD23$^{low}$ (right panel) B cells recovered from the second SHIP1$^{-/-}$ Ly5$^a$IgH$^b$ population when injected alone (left) or into mice injected 4 weeks before with a first Ly5$^b$IgH$^a$ population. Each bar represents the mean ± SD of values for 12-14 mice. Note that the number of activated Ly5$^a$ CD21$^{high}$CD23$^{low}$ and CD21$^{low}$CD23$^{low}$ present among the second set of B cells was not modified by the first set of B cells. **d,** IgM$^b$ levels produced by a second population of WT (left) or SHIP1-deficient (right) B cells transferred into Rag-deficient mice injected 4 weeks before with cells from WT IgH$^a$ donors expressed as a % of the control levels obtained after their transfer into naïve Rag-deficient hosts. Note that the IgM$^b$ production by SHIP1-deficient B cells was not inhibited in the presence of a first population of WT B cells. Each bar represents the mean ± SD of values for 12-14 mice. Data shown were pooled from two independent experiments. **e,** The left panel shows the IgM concentrations present in the serum of age-matched WT littermates and FcR$\gamma$IIB$^{-/-}$ mice. Each bar represents the mean ± SD of values for 8 mice. IgM titers in the mutant mice were significantly higher (** $p < 0.01$). The right panel shows the IgM concentrations present in the serum of age-matched WT littermates and SHIP1$^{-/-}$ mice. Each bar represents the mean ± SD of values for 4 mice. IgM titers in the mutant mice were significantly higher (* $p < 0.05$).

**Fig. 5** – (A) « Quorum sensing »: The presence of soluble IgG and the ability of the B cells to detect its levels are crucial to the homeostasis of the immune system. "Quorum sensing": The IgG secreted by activated B cells is detected (sensed) by the inhibitory Fc$\gamma$RIIB expressed by B cells that prevents further B cell activation. That is to say: overall B cell populations adapt

their behavior according to the sensing of the quantities of IgG produced. (B) Failure of "Quorum sensing" by defective sensor molecule: The inability to detect soluble IgG because of defects in the FcγRIIB expression (in FcγRIIB$^{-/-}$ mice) or signaling (in SHIP1$^{-/-}$ mice) leads to hyper IgM syndromes and autoimmune disease. (C) Failure of "Quorum sensing" by absence of the sensed molecule IgG also leads to hyper IgM syndromes and autoimmune pathology.

**Figure 1**

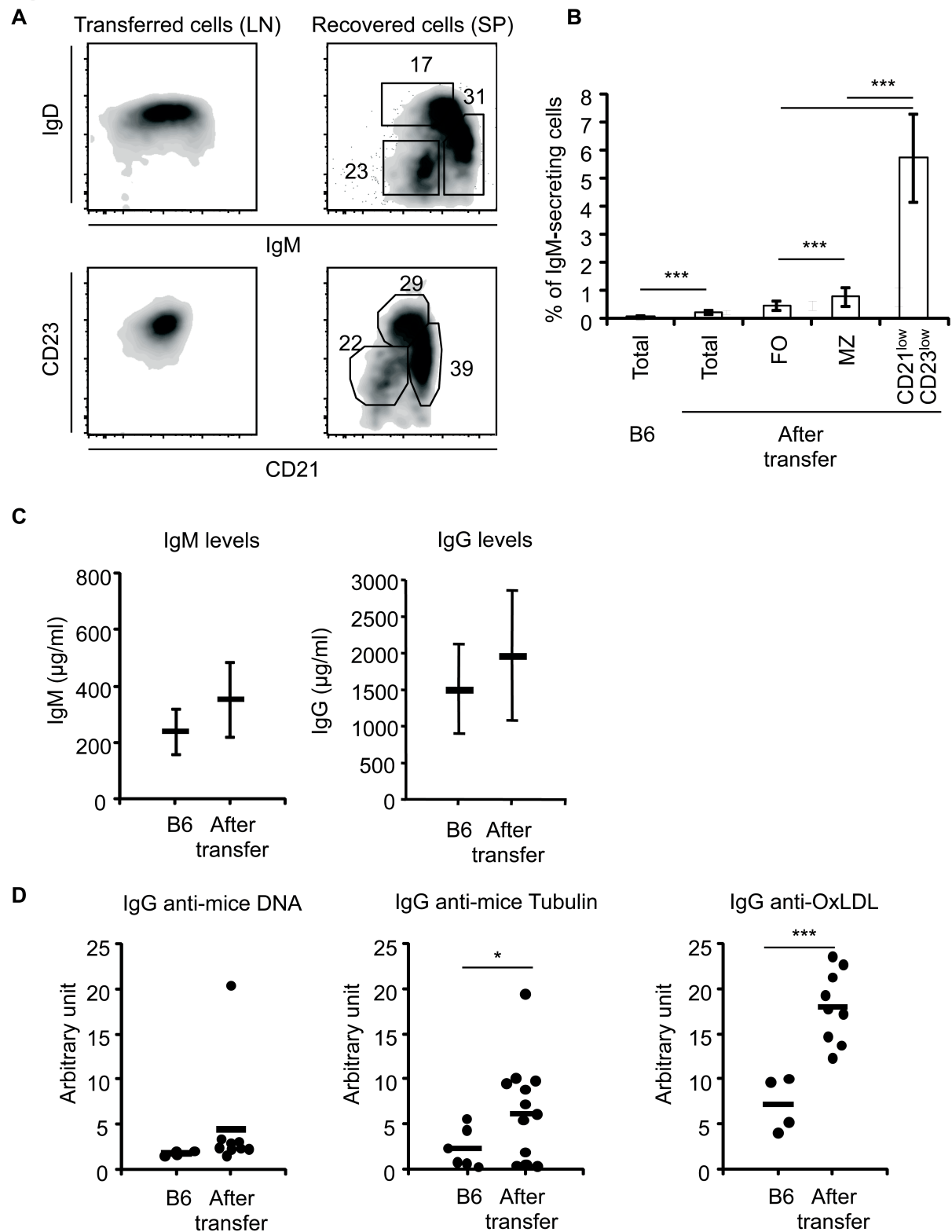

**Figure 2**

**A**

Number of Ly5$^b$ B cells from the 2$^{nd}$ population

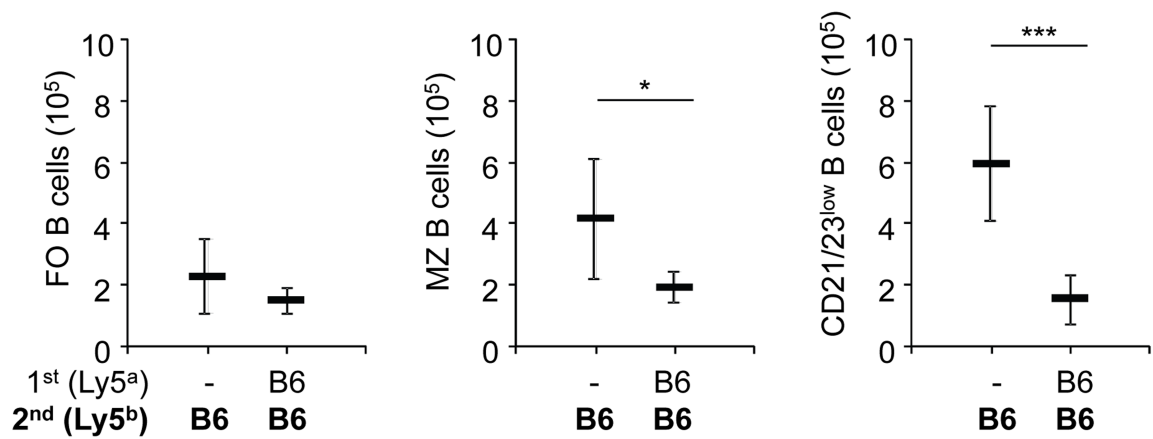

**B**

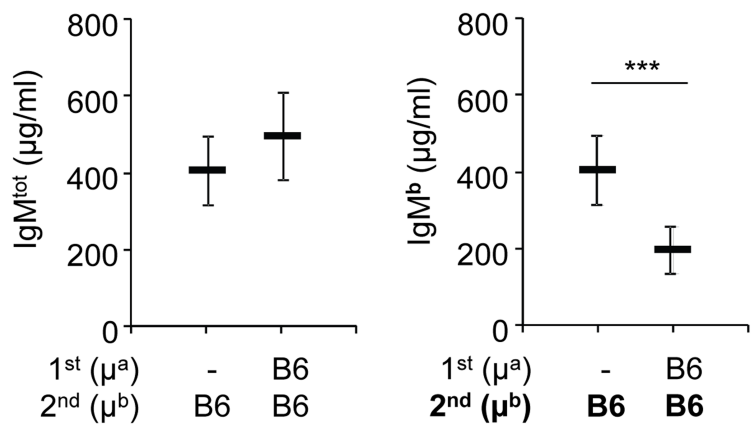

**Figure 3**

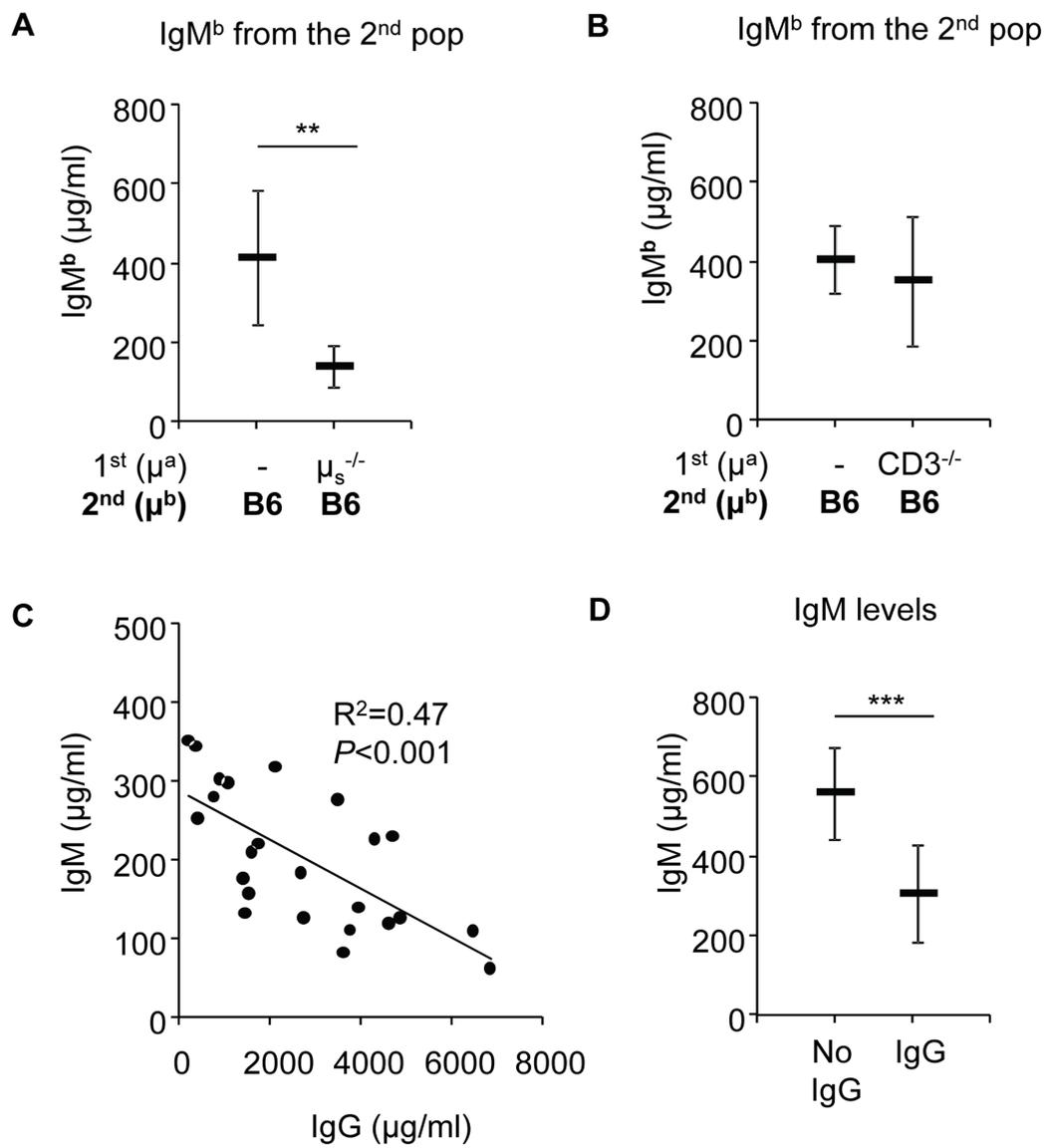

**A**  IgM$^b$ from the 2$^{nd}$ pop

**B**  IgM$^b$ from the 2$^{nd}$ pop

**C**

R$^2$=0.47
*P*<0.001

**D**  IgM levels

**Figure 4**

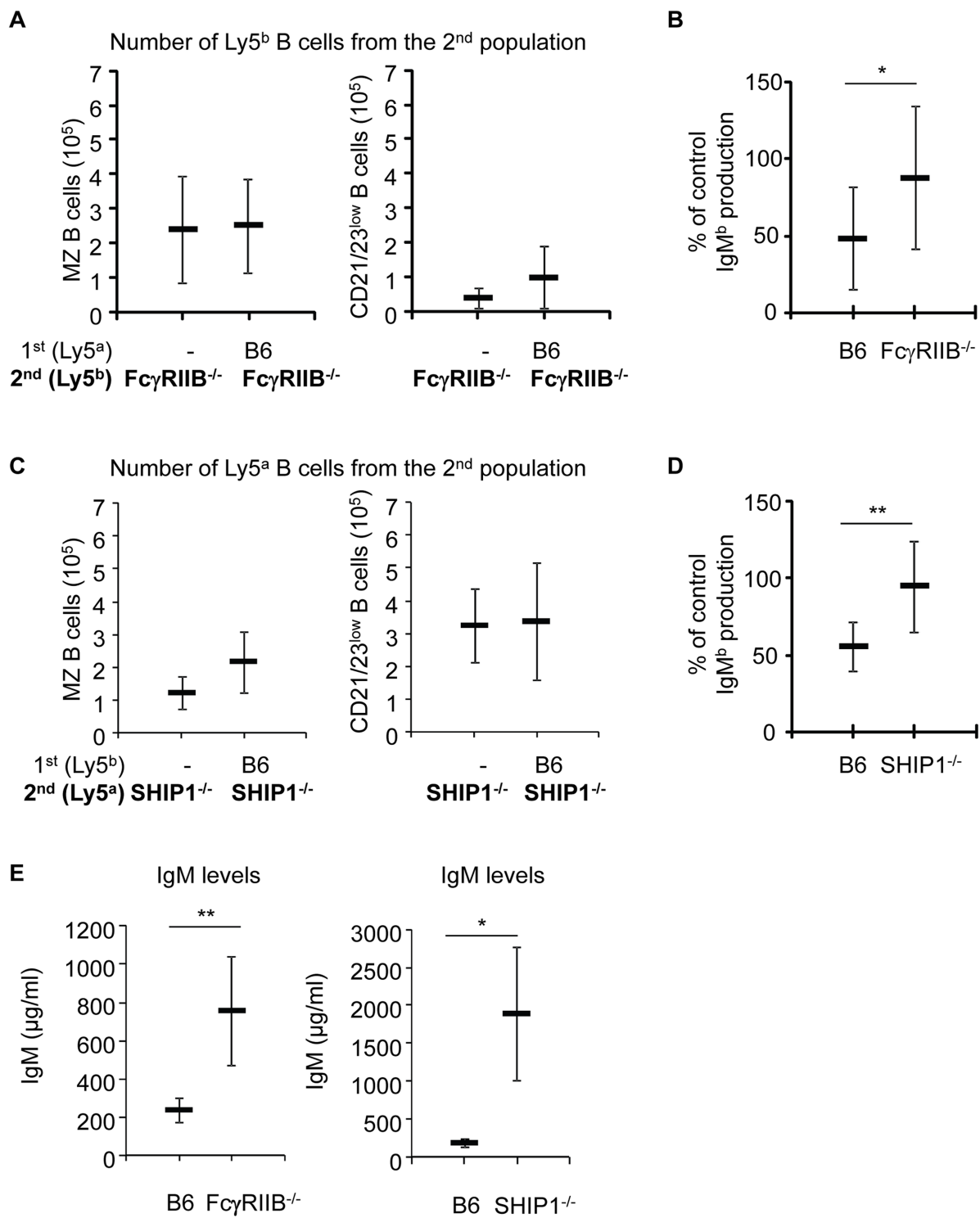

**Figure 5**

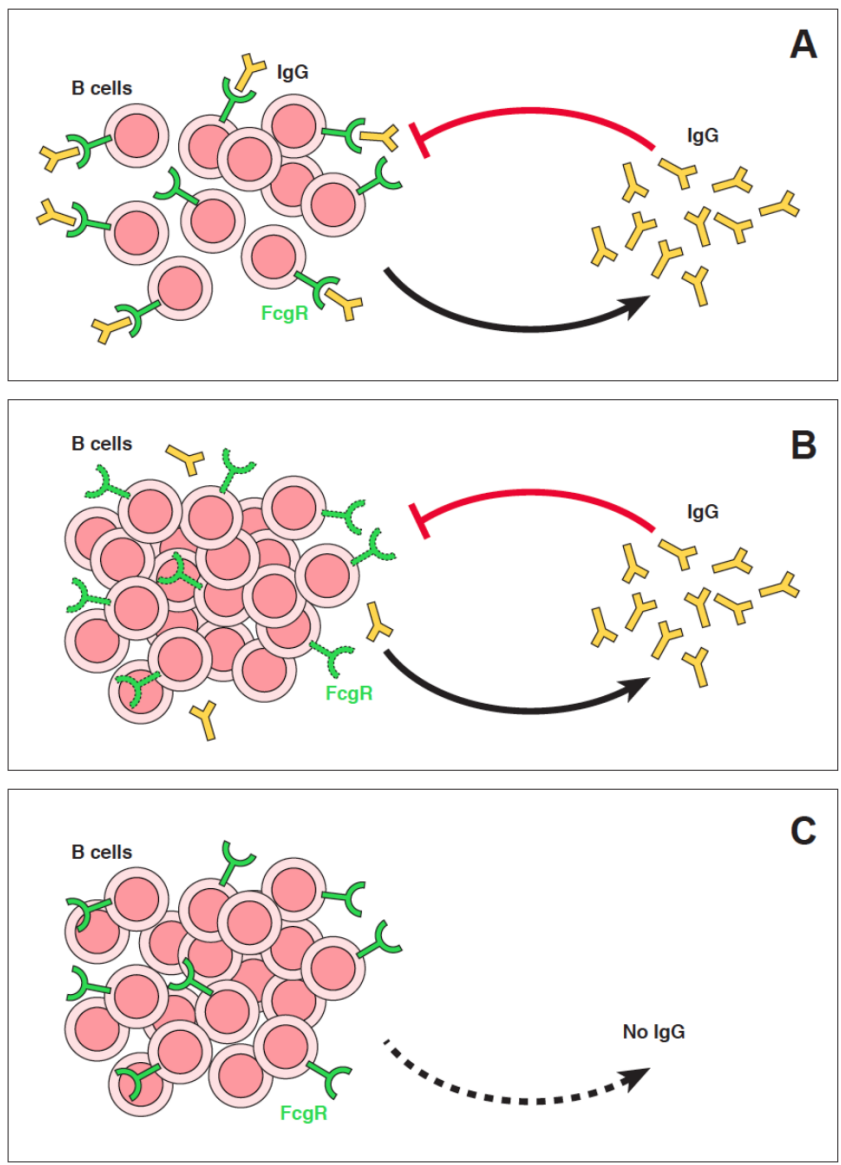